\documentclass[color]{aib}
\pdfoutput=1

\usepackage[hidelinks]{hyperref}
\usepackage{cite}
\DeclareGraphicsExtensions{.eps,.pdf,.jpeg,.png}

\usepackage{bytefield}
\usepackage[usenames,svgnames]{xcolor}
\newcommand{\colorbitbox}[3]{%
         \rlap{\bitbox{#2}{\color{#1}\rule{\width}{\height}}}%
         \bitbox{#2}{#3}}
\newcommand{\colorwordbox}[4][\empty]{%
         \rlap{\wordbox[#1]{#3}{\color{#2}\rule{\width}{\height}}}%
         \wordbox[#1]{#3}{#4}}

\usepackage[caption=false,font=footnotesize]{subfig}
\usepackage{fixltx2e}
\usepackage{url}
\usepackage{tikz}
\usepackage{amsmath}
\AIBtitle{{\rwthblue Support for Error Tolerance in the Real-Time Transport Protocol}}
\AIBauthors{Florian Schmidt, David Orlea, Klaus Wehrle}
\AIBnumber{2013}{19}
\AIBdate{December 2013}

\begin{document}

\makeAIBtitle

\title{Support for Error Tolerance in the Real-Time Transport Protocol\thanks{This technical report is an extended version of the conference paper \textit{A Heuristic Header Error Recovery Scheme for RTP}~\cite{rtp-refector} presented at IEEE/IFIP WONS 2013.}}
\author{Florian Schmidt, David Orlea, Klaus Wehrle}
\institute{Communication and Distributed Systems Group\\
RWTH Aachen University, Germany\\
Email: \{schmidt,wehrle\}@comsys.rwth-aachen.de, david.orlea@rwth-aachen.de}

\maketitle

\begin{abstract}
Streaming applications often tolerate bit errors in their received data well.
This is contrasted by the enforcement of correctness of the packet headers and payload by network protocols.
We investigate a solution for the Real-time Transport Protocol (RTP) that is tolerant to errors by accepting erroneous data.
It passes potentially corrupted stream data payloads to the codecs.
If errors occur in the header, our solution recovers from these by leveraging the known state and expected header values for each stream.
The solution is fully receiver-based and incrementally deployable, and as such requires neither support from the sender nor changes to the RTP specification.
Evaluations show that our header error recovery scheme can recover from almost all errors, with virtually no erroneous recoveries, up to bit error rates of about 10\%.
\end{abstract}

\section{Introduction}
Wireless communication is playing an ever-increasing role in Internet connectivity.
The ubiquity of notebooks, tablets, and smartphones leads to increasingly common use of at least one wireless hop for Internet communication, typically on the very last stretch from the user to an access point or base station.
One of the fundamental problems of wireless communication is the inherently higher unreliability of the link compared to wired communication, which leads to a higher bit error rate.
This high bit error rate, in turn, requires large numbers of retransmissions of data, as required by protocol standards that were defined with wired link characteristics in mind and require full packet retransmission if even a single bit error occurs.

At the same time, Internet traffic has seen composition shifts that lead to high volumes of video and audio traffic being transmitted.
Many of these codecs are in principle error-tolerant, being able to correct or at least mask errors.
However, especially for live streaming and bidirectional communication, they require high timeliness of data to reduce harmful delay.
For this class of traffic, partially erroneous packets arriving in time are helpful, while correct packets that arrive too late (potentially due to packet discards and retransmissions) are practically useless.
Therefore, providing those streaming applications with partially erroneous data is beneficial to their overall performance~\cite{hammer:corruptedspeech}.

It therefore stands to reason to allow applications that send and receive error-tolerant traffic to employ techniques that introduce error tolerance concepts into the standard Internet communication.
Previous solutions, most prominently UDP-Lite~\cite{rfc3828}, introduced error tolerance for a single protocol.
Their focus is typically on allowing \emph{payload error tolerance}, that is, they allow errors in the payload by using checksums to secure packet headers only.
For large payload sizes, for example, in video streaming, this works well because headers only form a small part of each packet.
Conversely, for communication such as Voice over IP (VoIP), packets are typically small, and the headers are, in relation, large, sometimes larger than the payload.
In such situations, employing a payload-only error tolerance is limited in its effectiveness.

We therefore focus on how to introduce \emph{tolerance also to header errors}.
We accept erroneous packets, even if the errors are within the header area, and heuristically repair these errors to identify the correct stream of data the packet belongs to.
In previous work~\cite{refector-conext}, we have shown that an approach that we termed \emph{Refector} (from Latin: repairer, mender) is feasible for UDP and IP, and that it can significantly reduce packet loss.
However, for many of our main application scenarios, UDP and IP alone are often not enough.
Many streaming applications, first and foremost VoIP, employ the application-layer Real-time Transport Protocol (RTP)~\cite{rfc3550} for timestamping, sequencing, and payload format layout.

The main contribution of this report is a heuristic header error recovery scheme designed for RTP that enables error-tolerant media codecs to receive packet payloads even if there are errors in the RTP header.
We identify which stream a packet belongs to by looking at the header values expected for the next packet in each stream, and then repair the header contents to those expected ones.
This way, we can repair errors for static parts of the header as well as dynamic parts that change from packet to packet.
Our system only needs to be deployed on the receiving side of a stream, does neither require any support from the sender nor change RTP's behavior, and as such is easily and incrementally deployable.
Recovery works well even at very high bit error rates up to 10\%, recovering most packets and almost never recovering incorrectly.
Our main envisioned application scenario is VoIP and audio conferencing, in which due to small payload sizes, header recovery will produce the highest relative gains over payload-only error tolerance solutions such as UDP-Lite, and in which many codecs support bit error tolerance.
However, the basic concept should also be applicable other scenarios that use RTP.

The rest of this report is structured as follows.
We discuss the design of our system, as well as the concept of heuristic header recovery, in Section~\ref{sec:design}.
In Section~\ref{sec:implementation}, we briefly describe our prototype implementation.
Evaluation results are presented in Section~\ref{sec:eval}.
We discuss related work in Section~\ref{sec:relwork} before concluding in Section~\ref{sec:conclusion}.

\section{System Design}
\label{sec:design}
In the following, we will first explain the concept of heuristic header recovery.
We will then give a short introduction into RTP, before we discuss the details of the recovery process for this protocol.

\subsection{Heuristic Header Error Recovery}
Our heuristic error tolerance scheme leverages the fact that at any given time, a protocol implementation has expectations about the contents of headers of received messages.
For example, RTP encapsulates media data into one or several so-called streams, each of which is assigned an identifier (the \textit{synchronization source identifier, SSRC}).
For every received packet, RTP requires the packet's header contents to match values expected for one of the streams.
If it cannot match the packet to any of them, the standard behavior is to discard it.

\begin{figure}
\centering
\includegraphics[width=0.7\columnwidth]{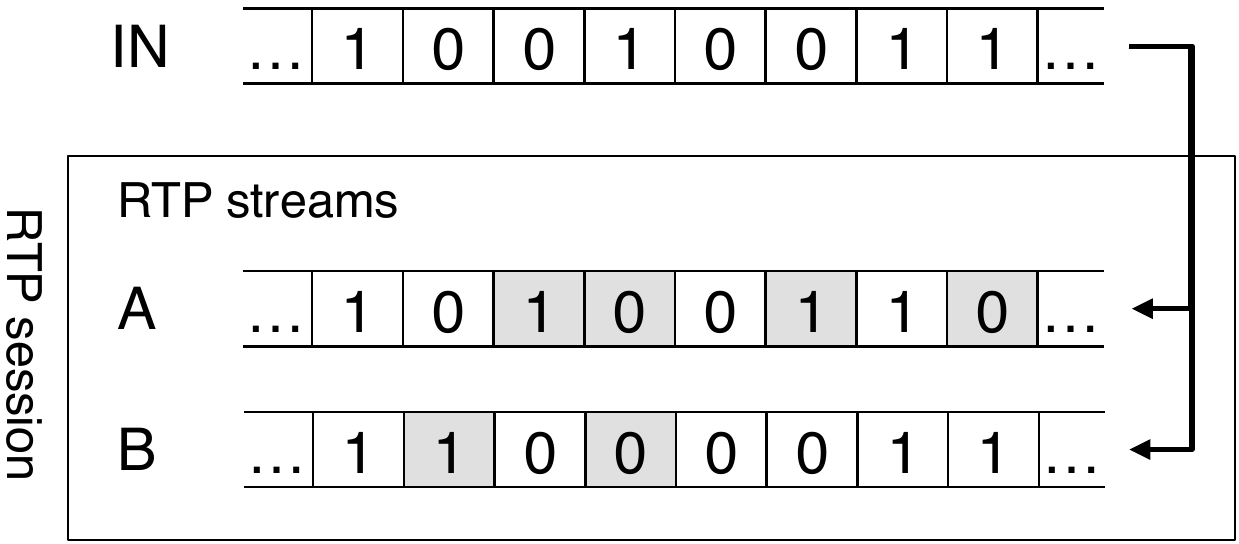}
\caption{A simple example of heuristic matching via Hamming distances. An incoming packet that contains errors is matched against two existing RTP streams. In this example, stream B is chosen because of the lower Hamming distance of 2 (as opposed to stream A with a Hamming distance of 4).}
\label{fig:hamming-decision}
\end{figure}

To heuristically repair header errors, as a first step, we do not outright discard packets as erroneous that contain unexpected header values.
Instead, we try to match a packet to the stream whose expected header values most closely match the received values.
As a similarity metric, we employ Hamming distances.
Hamming distances have several advantages: they are easy and computationally inexpensive to calculate, their similarity metric is independent of the position of the bit error in a string, and since bit errors are bit flips, they map to the amount of bit errors in a string very well.
A simple example is given in Figure~\ref{fig:hamming-decision}.

For the sake of clarity, only a small header excerpt is shown, and the remainder is assumed to match both streams perfectly.
In this case, the Hamming distance of the incoming packet is 2 to the expected values for the next packet of stream B and 4 to the expected values of stream A.
Therefore, the packet is assumed to belong to stream B, and assigned to it.

Because this matching strategy depends on heuristics, it can happen that it decides wrongly.
If a packet was corrupted in a way that makes it resemble a different stream's headers more closely than its own, it will be assigned to the wrong stream.
This problem of \textit{misattribution} is inherent to the system.
Since our solution focuses on error-tolerant media codecs, assigning data to the wrong stream is not immediately fatal (it will rather appear to such a codec as if it received a highly corrupted piece of payload); it is, however, undesirable, because the correct stream loses data, while another stream will have to cope with unrelated data.
Our main goal is therefore to maximize correct identification of endpoints, while reducing misattribution to very rare occurrences.

As the last step of the recovery process, the packet headers can then be repaired by replacing header values with those expected by the stream the packet was assigned to.
This is not strictly necessary, but generally advisable, because it will allow standard, unchanged protocol routines to process the packet properly.

\subsection{The Real-time Transport Protocol}
\label{sec:rtp}

The Real-Time Transport Protocol (RTP) is used for a wide range of audio and video streaming and conferencing scenarios.
For example, many VoIP solutions combine RTP as the streaming protocol with session protocols such as SIP~\cite{rfc3261} into a telephony system.

RTP is used in combination with a so-called profile that defines what media codecs can be used, and how they are encoded.
It can even, to a certain extent, modify the size and existence of header fields.
For this work, we will focus on the widespread profile that was standardized as baseline~\cite{rfc3551} together with RTP itself.
In this scenario, a telephony, conference, or other streaming setup comprises one or more RTP \textit{sessions}.
For example, a videoconference system is expected to use two sessions concurrently, one for video and one for audio, so that the two media types are separated from each other.

Each session comprises one or more \textit{streams} that identify logical units of data that belong together.
For example, each stream applies sequence numbers to packets that belong to it independently of other streams within the same session.
To identify streams within a session, each stream uses a \textit{synchronization source identifier} (SSRC) as unique originator ID.
Subsidiary contributors to a stream can be identified via \textit{contributing source identifiers} (CSRC).
The difference between SSRC and CSRC is not a technical, but a logical one.
Several cameras that show a scene might be considered different sources, such that each camera's stream uses its own SSRC.
A combined stream that uses some data from these cameras, on the other hand, might use its own SSRC and denote which camera the currently-sent data originated from by using that camera's CSRC.

Different RTP sessions are typically managed by different underlying protocol connections and will use different transport-layer ports.
This means that ``cross-talk'' between sessions can be ignored for the purposes of RTP error tolerance.\footnote{Identification of the correct receiver port, even under errors, is a different problem outside of the scope of this report; however, it has been shown~\cite{refector-conext} that such identification is feasible.}
Since packets of different sessions arrive on different ports, they will be easily distinguishable from each other.
In fact, an RTP library implementation will typically not even be aware of other sessions going on at the same time, because it will be independently instantiated once for each session.

The main focus of this report, our heuristic repair technique, therefore only has to address identifying streams correctly within a sessions, not sessions within a complete RTP setup.

\subsection{Header Field Categorization}

To support our header recovery scheme, we categorized the fields of the RTP header into three classes:

\emph{Static fields} are fields that are not expected to change in the lifetime of a stream.
They are either the same for all RTP packets (for example, the version field), or different, but immutable, for each stream (for example, the SSRC).
These fields are trivial to repair, because we know their possible value for each ongoing stream at any given point in time, and simply need to calculate the Hamming distance of the received header values to the static values of each stream. 

\emph{Predictably dynamic fields} change from packet to packet within a stream, but allow for prediction.
For example, a sequence number will predictably increase with every sent packet.
To repair these fields, we need to learn their behavior to predict the possible values.
For the sequence number, this is easy, because it will be incremented by 1 with every packet.
Hence, to match a received value to a stream, we need to match it against the next expected sequence number.

\emph{Unpredictably dynamic fields} change from packet to packet within a stream, and do so on an irregular basis.
These fields cannot be predicted, and therefore, errors in those fields are unrecoverable.
Note that this does not lead to outright drops of the packet, but rather to potentially incorrect values in those fields.

In the following, we will classify each of the RTP header fields into one of those categories.
Figure 2 gives an overview over the RTP header and an understanding of the proportions with which the three categories contribute to the header.
For those fields where it is applicable, we will explain the used recovery techniques.
(The techniques for fields that chiefly contribute to stream identification will instead be explained in Section~\ref{sec:streamid}.)

\begin{figure}
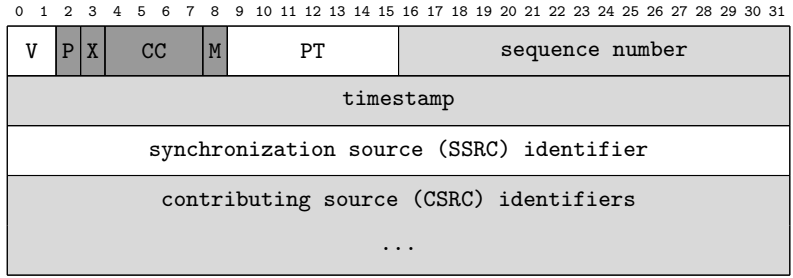

\centering
\definecolor{lgray}{gray}{0.85}
\definecolor{dgray}{gray}{0.6}
\definecolor{white}{gray}{1}
\begin{bytefield}[bitwidth=1em,boxformatting={\centering\small\ttfamily},bitformatting={\tiny\ttfamily}]{32}
\bitheader{0-31} \\
\colorbitbox{white}{2}{V} & \colorbitbox{dgray}{1}{P} & \colorbitbox{dgray}{1}{X} & \colorbitbox{dgray}{4}{CC} & \colorbitbox{dgray}{1}{M} & \colorbitbox{white}{7}{PT} & \colorbitbox{lgray}{16}{sequence number} \\
\colorbitbox{lgray}{32}{timestamp} \\
\colorbitbox{white}{32}{synchronization source (SSRC) identifier} \\ 
\colorwordbox[tlr]{lgray}{1}{contributing source (CSRC) identifiers} \\
\colorwordbox[blr]{lgray}{1}{...}
\end{bytefield}
\caption{A Real-time Transport Protocol header, with fields classified as static (white), predictably dynamic (light gray), and unpredictably dynamic (dark gray). Most fields are recoverable with our header recovery scheme; only seven bits are classified as unrecoverable, and even those are recoverable in many standard situations.}
\label{fig:rtp-header}
\end{figure}

\textit{Version (V)}: The version field contains a static identifier of the RTP version. Since up to now, there only exists one version, this field is always statically set to 2. During repair, this field can therefore be set to this value and disregarded otherwise.

\textit{Padding (P)}: This field signals whether the packet is padded after the payload.
Additional information, such as how much padding exists, is expected to be written into the padding itself.
We categorized this field as unpredictably dynamic, because padding may change from packet to packet in case of a combination of dynamic data rates and fixed block sizes expected by the media codec.
However, this combination would be rather exotic, and we deem that for most practical cases, this field will either be always 0 (either because the payload size is fixed and immutable, or because the media codec knows how to deal with varying payload sizes), or a fixed value for all frames.

\textit{Extension (X)}: If this field is set to 1, the RTP header is followed by an extension header that contains additional information as defined by a common sender and receiver profile.
Because choosing how and when to use extension headers is not defined by the standard, we categorized this field as unpredictable.
It should, however, be noted that the standard itself discourages the use of extension headers, and no currently defined official profile for RTP~\cite{rfc3551} makes use of them.
Therefore, for most practical cases, this field could be assumed to be a static 0.

\textit{CSRC count (CC)}: This field indicates the number of contributing source identifiers (CSRCs) that follow the SSRC in the RTP header.
If CSRCs are heavily used by the sender in numbers differing from packet to packet, this field is unpredictably dynamic.
That said, in a large number of RTP application scenarios, such as one-to-one VoIP or audioconferencing, CSRCs are of little use.
In these cases, each potential speaker would be identified by their own SSRC.
This is underlined by the fact support for CSRCs in many libraries and applications is rudimentary or non-existent, such as in linphone~\cite{url:linphone} and the oRTP library~\cite{url:ortp} that we use for our experiments.
In these cases, this field can assumed to be 0 and therefore trivially be repaired.

\textit{Marker (M)}: This bit is used to mark, in the words of the RTP standard~\cite{rfc3550} ``significant events''.
Which events are considered significant is up to the RTP profile that is used.
It can denote such diverse events as the beginning of a talkspurt (first packet to contain actual voice data after a comfort noise pause) or the last packet of a series of packets containing one video frame, if the data does not fit into a single packet.
Because of this diversity and dependency on the used profile, we categorized this field as unpredictable.

\textit{payload type (PT)}: The payload type signals the format of the payload for use by the media codec.
This only specifies the baseline codec, not specifics such as current bit rate.
As such, while it can be changed within the lifetime of a session, this is a rare occurrence.
For corrupted packets, we can therefore safely assume that the payload type is the same as seen in the last correctly received packet, effectively making it a static field.

\textit{sequence number}: The sequence number increases by 1 for every packet of a stream that is sent and is therefore predictably dynamic.
This is one of the fields that contributes to stream identification in a corrupted packet.

\textit{timestamp}: The timestamp indicates the point in time at which the recording of the payload contained in the RTP packet was started.
In audio situations such as VoIP and conferencing, this, as noted by the RFC, is expected to increase monotonously and regularly, as the RTP packet creation happens at regular intervals.
In these situations, timestamp and sequence number form a tightly coupled progression in that for every incrementation of the sequence number, the timestamp is increased by a static value, making it predictably dynamic.
This field also chiefly contributes to our way of stream identification.

\textit{synchronization source identifier (SSRC)}: As explained in Section~\ref{sec:rtp}, SSRCs identify streams within a session.
Because they are the identifier of a stream in RTP, they are static for each stream.
Together with the previous two fields, this forms the base of our identification for corrupted packets.

\textit{contributing source identifier(s) (CSRC(s))}: Also explained in Section~\ref{sec:rtp}, CSRCs identify different contributors to a stream.
These identifiers could give additional help in identifying streams.
This, however, depends on the correct reception of the CC field, because otherwise, CSRC headers that are part of the RTP header would be passed on to the application codec, or conversely, parts of the payload would be identified as CSRC.
As discussed for the CC field, due to the rarity of their use in main use cases of VoIP and audioconferencing, we can often safely assume that none are present in the received packets.

\subsection{RTP Stream Identification in Corrupted Packets}
\label{sec:streamid}
Generally speaking, every protocol that implements a form of multiplexing via header information needs to keep, for each communication end-point, information about the expected values in headers for each of these end-points.
In UDP, these are ports, while in RTP, this job is done by the SSRCs.

In the case of a corrupted packet that is received by the RTP protocol library, the most straightforward solution therefore is to check whether the SSRC matches one of the already ongoing connections.
If none of them matches perfectly, finding the closest match via Hamming distance might be the next step.
Still, in a case where the SSRC is strongly corrupted, this would be problematic.
However, it is possible to improve the matching by taking into account more fields for the overall matching decision.

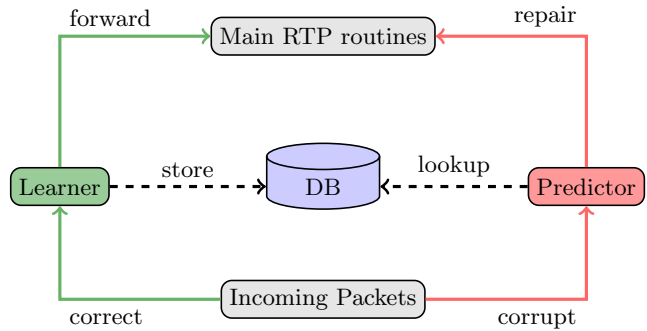
\begin{figure}
\center
\usetikzlibrary{shapes,arrows}
\begin{tikzpicture}[->,semithick]
\node[draw,
      fill=blue!20,
      shape=cylinder,
      aspect=0.5,
      minimum height=0.5cm,
      minimum width=1.5cm,
      shape border rotate=90] (data) at (0,0) {DB};
\node[draw,
      fill=Green!40,
      shape=rectangle,
      minimum height=0.5cm,
      minimum width=1.25cm,
      rounded corners] (learner) at (-3.5,0) {Learner};
\node[draw,
      fill=red!40,
      shape=rectangle,
      minimum height=0.5cm,
      minimum width=1.25cm,
      rounded corners] (predictor) at (+3.5,0) {Predictor};
\node[draw,
      fill=gray!20,
      shape=rectangle,
      minimum height=0.5cm,
      minimum width=2cm,
      rounded corners] (receive) at (0,-1.5) {Incoming Packets};
\node[draw,
      fill=gray!20,
      shape=rectangle,
      minimum height=0.5cm,
      minimum width=2cm,
      rounded corners] (forward) at (0,+2) {Main RTP routines};
\path[draw=black, dashed, very thick ] (learner) -- node[above] {store} (data);
\path[draw=black, dashed, very thick] (predictor) -- node[above] {lookup} (data);
\path[draw=Green!60, very thick] (learner) |- node[above right] {forward} (forward);
\path[draw=red!60, very thick] (predictor) |- node[above left] {repair} (forward);
\path[draw=Green!60, very thick] (receive) -| node[below right] {correct} (learner);
\path[draw=red!60, very thick] (receive) -| node[below left] {corrupt} (predictor);
\end{tikzpicture}
\caption{Packet flow through the extended RTP library. Correct packets will be inspected by the learner, their contents saved to a list indexed by SSRCs, and forwarded to the main RTP processing routines. Corrupted packets will be passed to the predictor, matched against expected header field values for ongoing streams, repaired to match the best stream, and afterwards forwarded.}
\label{fig:learnpred}
\end{figure}

Since each stream in RTP uses its own progression of sequence numbers and timestamps, these can be included into the overall decision.
To facilitate this, we learn additional state information from every correctly received packet to support us in the identification of corrupted packets.
The logical flow of packets through our learner--predictor setup is shown in Figure~\ref{fig:learnpred}.

Whenever a correct packet is received, the \textit{learner} saves the header contents for future use.
For each ongoing stream, it saves the last correctly received packet in this fashion.
Furthermore, it calculates the sampling rate, that is, the difference between two consecutive (as identified by the sequence number) packets in their timestamps.
That way, whenever a corrupted packet is received, the second component, the \textit{predictor}, can match the header field contents by matching SSRCs, the received sequence number against the recorded sequence number incremented by 1, and the received timestamp by the recorded timestamp incremented by the sampling rate.
This means that a much larger area of the header can be compared to expected values, and the finding of a best match via Hamming distances becomes inherently more stable.

This will work well unless more than one corrupted packet is received in sequence.
In that case, a simple incrementation by 1 resp. the sampling rate is not effective, because any further packets will not have the chance to exactly match this information.
One solution to this would be to have the learner update the information for use by the predictor with the header contents of the received corrupted packets.
This is risky, because the learner would save potentially broken information for future use.
Therefore, we instead save, for every stream, a \textit{bad packet counter} that tracks how many corrupted packets were assigned to that stream since the last correct packet was received for it.
We then use this value as a multiplier to the increments for sequence number and timestamp.
While completely lost packets or those misattributed to the wrong stream will still cause a slight desynchronization between received and expected information, a number of corrupted packets received in succession will not.
Whenever the learner receives a correct packet for a stream, it is able to update all header information, specifically the sequence number and timestamp, from that packet's headers, and therefore resets the bad packet counter.

\subsection{User--Kernel interface}
As explained in Section~\ref{sec:streamid}, our approach learns from correct packets to correctly identify corrupted packets with errors in header fields.
Because RTP does not employ any checksumming, instead relying on error-detection by lower layers, it is not able to reliably detect errors on its own.
Furthermore22, to even receive erroneous packets in the first place, it will need to instruct the operating system's network stack to pass those to it.
We therefore extend the standard socket interface for user-space--kernel-space interaction in a similar way to~\cite{refector-conext}.

In our Linux implementation, we extend the \texttt{recvmsg} system call that is used to receive incoming packets and that also signals ancillary information with an additional flag \texttt{MSG\_HASERRORS} that signals whether checksums at lower layers failed.
Note that this is a conservative check: any RTP packet that contains errors will have checksum fails (barring extremely rare circumstances of undetectable errors that are inherent in every checksumming system and not specific to our system) reported.
Conversely, if errors solely occurred in headers of lower layers, the potential for errors in the RTP packet will be signaled when in fact, there are no errors present.
This means that we can always rely on the fact that the learner will not learn from corrupted packets, and our per-stream information (see Section~\ref{sec:streamid}) will be reliable.

To signal to the network stack that our RTP library is able to cope with errors and interested in erroneous packets, we added an additional socket option \texttt{SO\_BROKENOK} that can be set by an application for every socket.
That way, the network stack can still discard packets for legacy and error-sensitive applications (for example, concurrent file transfers), making reception of erroneous packets and opt-in choice.

Of course, to facilitate this reception of erroneous packets, changes have to be made to the network stack's error handling.
For the scope of this report, we abstract from this problem and assume that a solution such as~\cite{refector-conext} is in place.

\section{Implementation}
\label{sec:implementation}
For this work and to evaluate our concepts, we implemented the learner--predictor scheme for heuristic header error recovery into the oRTP~\cite{url:ortp} library (version 0.16.5), an open-source library that was easily adaptable for our purposes.

The implementation follows a minimally invasive approach that interferes with the standard behavior of the RTP packet handling as little as possible.
This is advantageous for such tasks as statistics collection that can be used to inform the sender about current reception conditions via RTCP~\cite{rfc3550} so it can potentially decide on proper reactions to improve streaming quality.

Whenever a correct (error-free) packet is received, the learner takes the header of the RTP packet and saves it to its list of current streams, indexed by SSRC, so that for every stream, only the most recent header is saved.
In addition, it calculates the sampling rate between the just received and the last saved packet by $$\frac{ts_{this} - ts_{last}}{seq_{this} - seq_{last}}$$
where \textit{ts} and \textit{seq} denote the values of the timestamp and the sequence number fields, respectively.

Whenever a packet flagged as erroneous is received, the predictor matches the received header to the possible expected headers of each current stream, iterating over the list maintained by the learner.
During this matching, each of the expected headers will have their sequence number field increased by the bad packet counter (see Section~\ref{sec:streamid}), and the timestamp by that value multiplied with the sampling rate calculated by the learner.

Whenever the predictor has decided on which stream the packet most likely belongs to, it will attempt header repair by copying the saved header from the predictor's list over the received header, updating the sequence number and timestamp value accordingly.
This way, the subsequent RTP routines implemented in the oRTP library do not have to be changed to introduce additional error handling, since the header is now guaranteed to be coherent.
Finally, it will increment the bad packet counter of that stream.

The ease of this approach and the little need for in-depth changes in the oRTP library suggest that similar changes in other RTP implementations should be similarly easy and fast to implement.

\section{Evaluation}
\label{sec:eval}
Since the main advantage of our heuristic header error recovery scheme is that erroneous packets can be delivered to a stream, our evaluation focuses on two packet-delivery-related metrics:
(1)~How often can a packet be delivered to the correct stream?
And (2)~how often does the heuristic approach misidentify the stream the packet belongs to, and misattributes it to the wrong stream?
To answer these questions, we will present several setups that we evaluated. We will start with a description of the evaluation setup before discussing results.

\subsection{Experimental Setup}
\label{sec:evalsetup}
We implemented our heuristic header repair into the oRTP library, version 0.16.5, and conducted our evaluation on a Ubuntu Linux 10.04.

To specifically evaluate the behavior of our heuristic header repair system for RTP, we wanted to eliminate influences from the lower layers that can skew our evaluation results.
To ensure this, we exchanged the standard network socket interface that the RTP library uses to open connections via the network stack with Unix Domain Sockets.
These allow data exchange between processes in very similar way to network communication, without additional protocols being used.
In effect, the RTP packets could be exchanged between two instances of the library running on the same machine, without changing the behavior of the implementation.

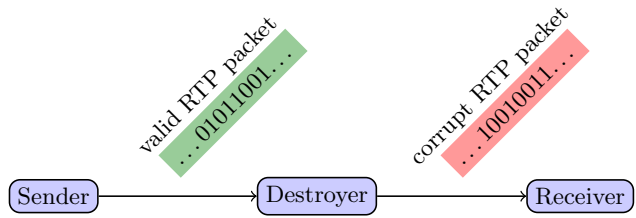
\begin{figure}
\center
\usetikzlibrary{arrows,shapes}
\begin{tikzpicture}[->,semithick]
\node[draw, fill=blue!20, shape=rectangle, rounded corners, minimum height=1.5, minimum width=0.5] (sender) at (0,0) {Sender};
\node[draw, fill=blue!20, shape=rectangle, rounded corners, minimum height=1.5, minimum width=0.5] (destroyer) at (3.5,0) {Destroyer};
\node[draw, fill=blue!20, shape=rectangle, rounded corners, minimum height=1.5, minimum width=0.5] (receiver) at (7,0) {Receiver};
\path[draw=black] (sender) -- (destroyer);
\path[draw=black] (destroyer) -- (receiver);
\node[fill=Green!40, shape=rectangle, rotate around={45:(-1.5,-0.25)}] at (3cm,0.25cm) {\dots01011001\dots};
\node[fill=red!40, shape=rectangle, rotate around={45:(-1.5,-0.25)}] at (6.75cm,0.25cm) {\dots10010011\dots};
\node[rotate around={45:(-1.5,-1)}] at (3.2cm,0.7cm) {valid RTP packet};
\node[rotate around={45:(-1.5,-1)}] at (6.95cm,0.7cm) {corrupt RTP packet};
\end{tikzpicture}
\caption{Overview over the experimental evaluation setup. The sender and receiver are two instances of the RTP library. The receiver is enhanced with our heuristic header error recovery techniques. The destroyer can be instructed to introduce various amounts of errors into the RTP packet.}
\label{fig:evalsetup}
\end{figure}

To introduce errors into the RTP packets sent by the sender instance, we did not connect the two instances directly.
Instead, we created a packet destroyer, whose sole job is to introduce bit errors into a data stream with a defined probability $p$.
For each bit, it rolls a random number between 0 and 1, and flips the bit if the number is lower than $p$.
This, in effect, implements a Bernoulli process.
The evaluation setup is depicted in Figure~\ref{fig:evalsetup}.

For each step in our evaluation, we investigated three scenarios that differed in the number of concurrent streams in the RTP session.
While a single-stream session is arguably the most common use case for RTP, especially in the case of VoIP telephony, this case is also not very interesting from an evaluation point of view.
In fact, in a single-stream scenario, since there is no risk of misattribution to other concurrent streams, our repair technique will be able to correctly assign \textit{every} packet, regardless of error rate.
To investigate the possible downsides and limits of this approach, we therefore looked at scenarios with several concurrent streams, namely two, three, and four concurrent streams.
An experiment was defined by the combination of the number of concurrent streams and the bit error rate.
The sender sent 10\,000 packets for each stream in every experiment.
In each experiment, the first two packets were not corrupted in any way, to populate the list of known SSRCs.
This can be considered to model a use case in which initial setup of an RTP connection was successful and link quality started to suffer afterwards.
Experiments were repeated 10 times for every data point unless otherwise noted.
This is especially important since oRTP follows the RFC advice and randomizes SSRCs, as well as initial timestamps and sequence numbers, every time a stream is created.
This influences the robustness of the heuristics depending on how large the Hamming distances between values of different streams are.
Error bars in graphs denote 95\% confidence intervals.
In some cases (Figures~\ref{fig:hd-cutoff-misatt} and~\ref{fig:hd-cutoff-drop}), error bars were not plotted to preserve lucidity.

\subsection{Misattribution}

As a first step, we ran our experiments for two, three, and four concurrent streams in an RTP session.
In this setup, every packet was assigned to the stream it most likely belonged to.
That is, the (potentially erroneous) header was assigned to the stream whose expected header values it had the lowest Hamming distance to.
Every packet was assigned to some stream, without discarding any packets.
The main downside of this is that it is possible for a header to be corrupted in a way that it more closely resembles another stream's header on the receiver's side.
In this case, the packet will be misattributed.

\begin{figure}
\centering
\includegraphics{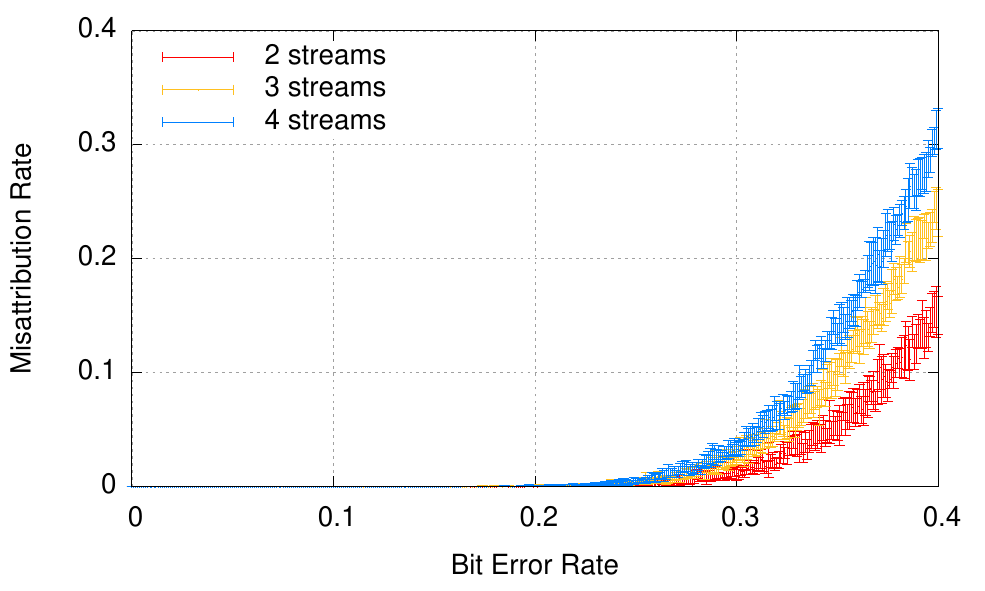}
\caption{Misattribution rates for two, three, and four concurrent streams with no cutoff (i.e., the best match is always taken, even if its Hamming distance is large; packets are never dropped). For each bit error rate (increments of 0.001 from 0 to 0.5), the mean and 95\% confidence intervals are shown. Even at a bit error rate of 20\%, we witnessed virtually no misattributions.}
\label{fig:ber-independent-nocutoff}
\end{figure}

The results from this experiment are shown in Figure~\ref{fig:ber-independent-nocutoff}. As expected, this risk increases with high bit error rates and the number of concurrent streams.
High bit error rates lead to more corruption in the header; as an extreme case, at 50\% BER, a header can be expected to be a random bit sequence without any resemblance to its original contents.
High bit error rates therefore make it harder to recover and assign the packet to the correct stream.
As the number of concurrent streams in an RTP session increases, on average the Hamming distances between between streams will decrease, making it harder to distinguish them.
To correctly recover and assign the packet, its header values must be closer to the expected header values of the correct stream than those of any other stream.

One important finding to point out here is that our heuristic error recovery scheme does produce almost no misattributions until the BER is in excess of 20\%.
For comparison, BERs of more than 0.1\% can lead to a packet drop rate of almost 100\% in standard systems, and even if data reaches the media codec, most voice codecs will start showing noticeable degradation at that BER~\cite{nguyen:spacevoip}.

\subsection{Field Errors}

\begin{figure}
\centering
\includegraphics{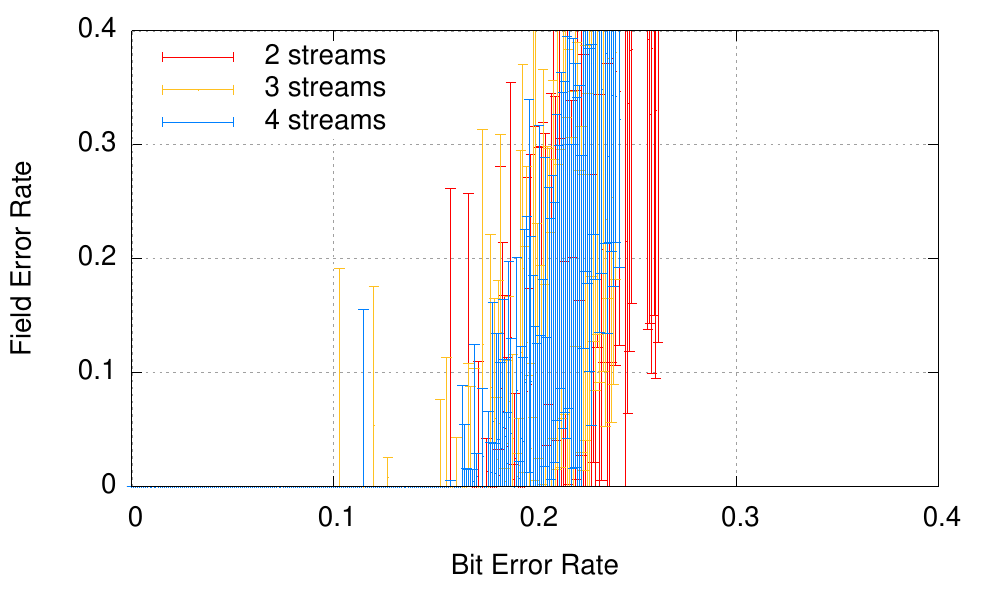}
\caption{Errors in header fields due to incorrect repairing. The high variance is due to error propagation that leads to incorrect repair of header fields in a large number of packets after a single misattribution, especially at high BERs.}
\label{fig:ber-fielderrors}
\end{figure}

Misattribution is only one type of error that can occur in heuristic recovery.
The other type occurs if packets are not only assigned to a stream, but also repaired to the values that this stream expects in the next packet's header.
In this case, fields can be wrongly repaired.
The occurrences of these \textit{field errors} are shown in Figure~\ref{fig:ber-fielderrors}.

The high error rates and very large uncertainties denoted by the confidence intervals originate in the fact that field errors are prone to error propagation.
As an example, consider two concurrent streams $A$ and $B$ that expect as next sequence number $s_A$ and $s_B$, respectively.
If packet $n$ belongs to stream $A$, but is misattributed to $B$, its sequence number will be incorrectly repaired to $s_B$.
In addition, packet $n+1$ will now also suffer from incorrect repair of the sequence number, regardless of whether it belongs to stream $A$ or $B$, and even if it is assigned correctly.
If it belongs to $A$, it contained $s_A+1$, which will be repaired to $s_A$.
If it belongs to $B$, it contained $s_B$, which will be repaired to $s_B+1$.
This shift by 1 will continue until the stream ``resynchronizes'' after the reception of and learning from a correct packet.
At high bit error rates, almost every packet will be corrupted, so that it can take a long time until this error is corrected.

Two facts are of note here:
(1)~The field error rate overestimates the impact that such an error has.
In the sequence number case, while a large number of them might be incorrect, they are simply shifted by a fixed number.
Interruptions in the regular pattern only occur at the time of misattribution and resynchronization.
This is the only point in time playback would be negatively affected by this problem.
(2)~Because these errors are a secondary effect of misattributions, they do not occur at BERs below 10\%.
Again, this is much higher than the typically tolerable BER for media transmissions.

\subsection{Reduction of Misattribution}
\label{sec:cutoff}

As shown in the last two sections, our heuristic header error recovery can be expected to work exceedingly well in most situations, correctly assigning all packets to their corresponding streams and producing no incorrect repair.
One noticeable exception to this rule is that in cases such as short-term interference, the RTP header could experience a much higher BER than the surrounding areas (lower-level protocols and payload).
In this case, RTP packets could be received without problems by the RTP library, and the payload could still be useful for the media codec, but the high header BER can lead to misattributions.

We therefore investigated when misattributions tended to happen.
As a self-imposed limitation, we only used information that was available to the RTP library itself, so that our findings could then be used to improve the performance of our implementation.
One information that is available, after comparing the received header to the expected header values of all streams, is the Hamming distance to the best match.
Low Hamming distances mean a very close match, while high distances mean only rough resemblance.
That means that the probability that a misattribution occurs should be higher whenever a high Hamming distance to the closest match is observed (because it is relatively unlikely that random bit errors corrupt a header in such a way that it then happens to exactly match another stream's expected header).

\begin{figure}
\centering
\includegraphics{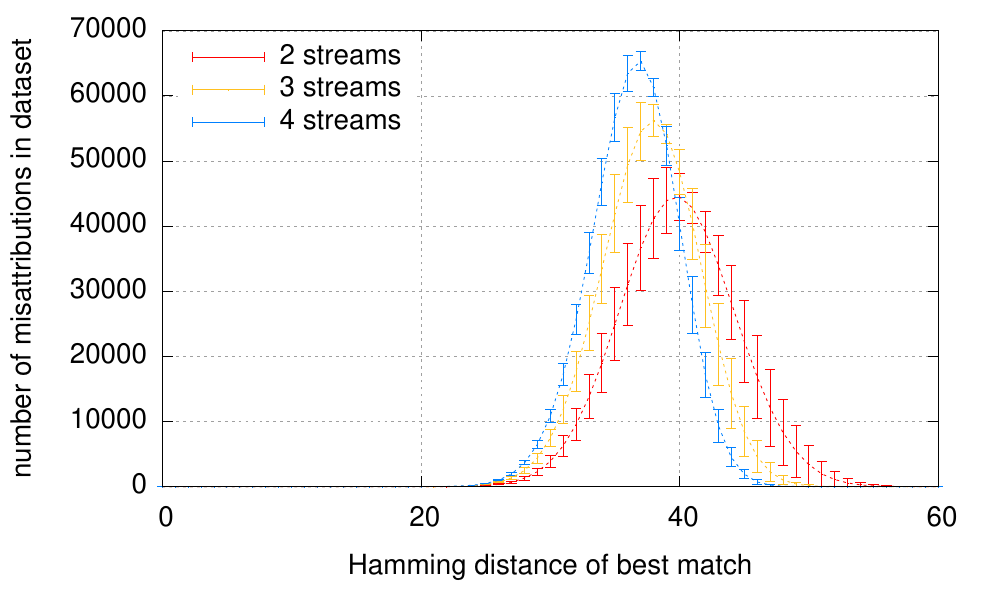}
\caption{Misattribution occurs due to high bit error rates. A side effect of high BERs is that even the best match often shows a high Hamming distance to its expected header values. Hamming distance can therefore be used as an estimator for risk of misattribution. At low Hamming distances, no misattribution occurs. Extremely high Hamming distances are uncommon because it is more likely another match with lower distance exists instead.}
\label{fig:hd-misatt}
\end{figure}

The results of this investigation are shown in Figure~\ref{fig:hd-misatt}.
Indeed, misattributions are virtually nonexistent at Hamming distances of less than 20.
For comparison and reference, a minimum RTP header without any CSRC fields has 12 bytes, that is, 96 bits.
20 bits therefore translates into more than 20\% BER, which is consistent with our results from Figure~\ref{fig:ber-independent-nocutoff}, in which we witnessed almost no misattributions until that BER.
Note that Figure~\ref{fig:hd-misatt} shows the absolute number of misattribution in our experiments, which is why the misattribution number decreases again at high Hamming distances since these are less likely to occur.

\begin{figure}
\centering
\includegraphics{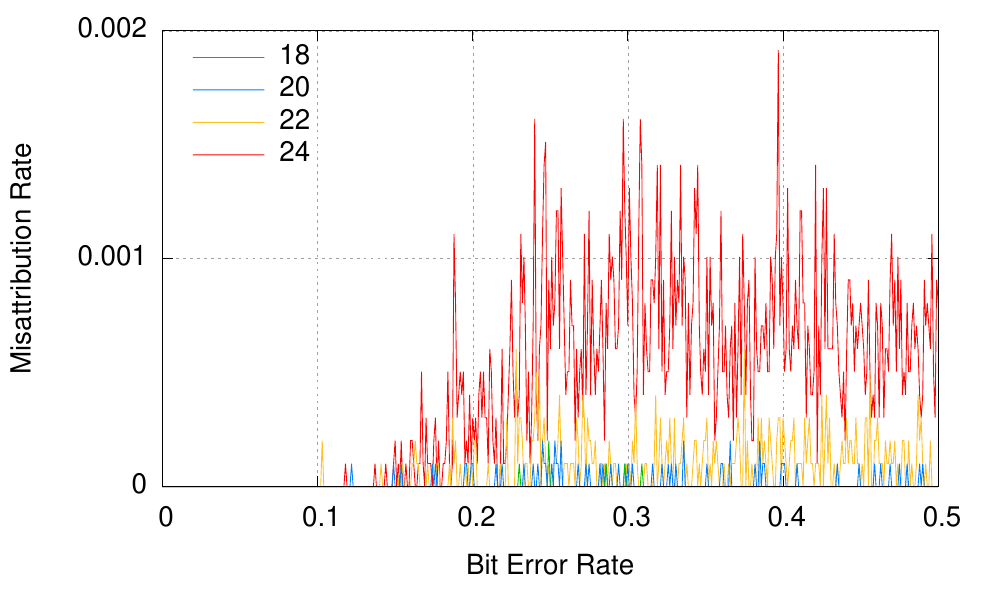}
\caption{Misattribution rate in a 4-stream scenario for different Hamming distance cutoffs. By setting a a cutoff and discarding all packets whose best match has a higher distance, misattribution can be reduced dramatically. Even at a cutoff value of 24 (one quarter of the 96 bits of the standard RTP header), misattribution stays below 0.2\%, regardless of BER, and becomes exceedingly rare at stricter cutoffs.}
\label{fig:hd-cutoff-misatt}
\end{figure}

Considering this result, we then decided to change our implementation by having it drop all packets that show a high Hamming distance to the best match.
The goal is to reduce misattributions at high BERs, instead dropping those suspicious packets.
Figure~\ref{fig:hd-cutoff-misatt} shows misattribution rates for four different Hamming distance cutoffs in a 4-stream scenario.
We focus on the 4-stream scenario here because it showed the highest misattribution rate in our initial experiments (cf.~Figure~\ref{fig:ber-independent-nocutoff}).
The results show that this cutoff is very effective in reducing misattributions.
Even at the more lenient cutoff of 24, misattributions are reduced by almost two orders of a magnitude; cutoffs of 20 and, even more so, 18 almost completely eliminate misattribution.

However, as pointed out above, this improvement comes at a cost: as a tradeoff, we will now drop packets with high Hamming distances to their best match, even if they would have been assigned to the correct stream.
We therefore again evaluated the 4-stream scenario for the four chosen cutoff rates, this time investigating the average packet drop rate at different BERs.

\begin{figure}
\centering
\includegraphics{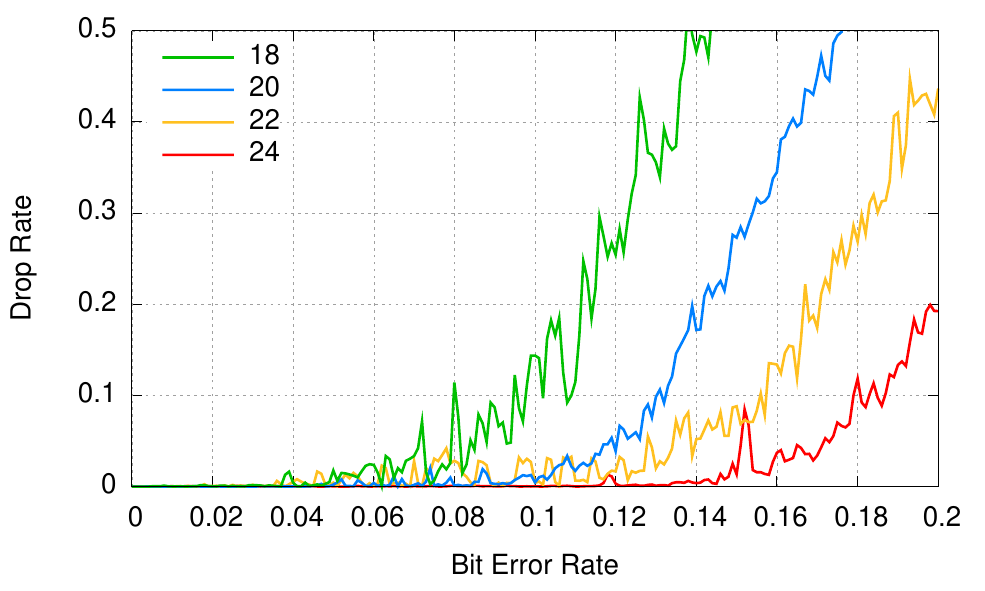}
\caption{Drop rate in a 4-stream scenario for different Hamming distance cutoffs. When using cutoffs, misattribution rate is reduced, but packets are outright dropped instead. However, even at a strict cutoff rate that prevents virtually all misattribution (see Figure\,\ref{fig:hd-cutoff-misatt}), drop rates stay minimal until BER is in excess of 5\%.}
\label{fig:hd-cutoff-drop}
\end{figure}

Figure~\ref{fig:hd-cutoff-drop} shows that we indeed considerably increased the drop rate compared to the misattribution rate of Figure~\ref{fig:ber-independent-nocutoff}.
For the strictest cutoff value of 18, drops start to occur at about 4\% BER, and reach 10\% drop rate at 10\% BER. 
The less strict cutoff of 20 only has a a drop rate of about 1\% at 10\% BER, and the most lenient cutoff of 24 rarely drops any packets until the BER is in excess of 15\%.

Judging from these results, we suggest to set the cutoff to 20 bits as a good tradeoff between prevention of both misattribution and packet drops.
This should effectively prevent most misattributions in most situations (single-digit number of concurrent streams, independently of bit error rate), while still only regularly dropping packets when the BER reaches excessively high values above 10\%.

\section{Related work}
\label{sec:relwork}
While the concept itself of heuristically recovering from header errors is relatively unique, related work can be roughly separated into two fields: those that try to optimize the reception and retransmission of data (typically residing on the PHY and MAC layer), and those that ignore errors in the payload of data packets.

Considering the first group, one solution proposed is to improve ARQ (automatic repeat request, i.e., (N)ACK-based mechanisms with retransmissions) by partitioning packets and calculating and sending partial checksums instead of a checksum that covers the whole packet.
Maranello~\cite{han:maranello} uses this idea so that the receiver can identify the corrupt blocks and selectively request retransmissions of those.
By using rich data from the physical layer, PPR~\cite{jamieson:PPR} uses so-called soft information, that is, additional information from the signal demodulation and decoding unit about the probability that a bit is 1 or 0, to recognize erroneous parts in packets and selectively retransmit those parts.
The same kind of rich information is used to reconstruct a correct packet from several receptions, either due to spatial diversity, or from retransmissions of the same packet.
Examples include SOFT~\cite{woo:soft} (spatial diversity), ZigZag~\cite{gollakotta:zigzag} (retransmissions), and MRD~\cite{miu:mrd} (both).
All of these approaches either require deployment on all participating nodes to be effective, or special hardware that allows access to physical layer soft information, or both.

The second group, using checksums with partial checksums to only secure headers or parts of the payload, is most prominently represented by UDP-Lite~\cite{rfc3828}, an RFC-standardized protocol derived from UDP by redefining the ``length'' header field as ``checksum coverage''.
While the protocol is easily implemented, and, once available, the switch to UDP-Lite from a UDP-based application is simple, there are several drawbacks.
Both sides need to be able to understand UDP-Lite, as it is, in effect, a new transport layer protocol.
This also means that application developers have to rely on the operating system to provide the protocol, since they cannot provide it themselves.
UDP-Liter~\cite{lam:udp-liter} solves these problems by providing a way for applications to receive packets with UDP checksum errors.
It ignores checksum mismatches and passes the payload to the application.
However, UDP-Liter does not provide mechanisms to recover from header bit errors.
Packets with errors in the UDP header are lost, as are all packets with errors on lower layers because UDP-Liter focuses solely on UDP.

With respect to heuristic recovery, Jiang~\cite{jiang:errortolerance} and Schmidt et al.~\cite{refector-conext} proposed solutions to heuristically recover from header bit errors.
Both of these chiefly rely on Hamming distances as similarity metric to some pool of expected values.
Jiang~\cite{jiang:errortolerance} focuses on header recovery for some static fields in the 802.11 MAC header, such as MAC addresses, as opposed to sequence number fields, which are not considered.
Schmidt et al.~\cite{refector-conext} aim at recovering headers in the IP and UDP protocol.
Again, the focus is on the static fields that form the bulk of information in those protocol headers.
One difference is that we propose a simple but effective way to deal with dynamic header fields that follow regular patterns, such as the RTP sequence number and timestamp field.

\section{Conclusion}
\label{sec:conclusion}
In this report, we presented a heuristic header error recovery scheme for RTP that allows identifying the stream a packet belongs to even if errors in header fields occurred and repairing those fields.
We showed that our scheme is robust up to bit error rates of 10\%, very rarely assigning packets to wrong streams or repairing incorrectly, while still keeping packet drop rates low.
This holds true even more at more modest BERs that are realistically tolerable by media codecs.

One main field of future work is to make the statically defined Hamming cutoffs described in Section~\ref{sec:cutoff} dynamic, adapting to both the number of concurrent streams and the Hamming distance in SSRC, timestamp, and sequence number of those streams.
That way, the tradeoff between reducing misattribution and increasing drop rate can be further optimized.
Another idea would be to loosen the requirement to have a fully receiver-based solution.
If the sender is also adapted for heuristic header error recovery, it can choose, for concurrent streams, SSRCs and initial values for timestamps and sequence numbers that maximize the Hamming distance, increasing robustness.

Overall, we consider the work presented in this report a feasible, simple, and effective approach to support an error-tolerant application in the reception of partially erroneous data.

\section*{Acknowledgments}
This research was funded in part by the DFG Cluster of Excellence on Ultra High-Speed Mobile Information and Communication (UMIC).

\cleardoublepage
\section*{Aachener Informatik-Berichte}
\newfont{\sss}{cmr10 scaled 1000}
\newfont{\bbb}{cmbx10 scaled 1000}
\sss

{\bbb This list contains all technical reports published
  during the past three years.
  A complete list of reports dating back to 1987 is available from:
\begin{center}
  \url{http://aib.informatik.rwth-aachen.de/}
\end{center}
  To obtain copies please consult the above URL or send your request
  to:
\begin{center}
  Informatik-Bibliothek, RWTH Aachen, Ahornstr.~55, 52056 Aachen,\\
  Email: \email{biblio@informatik.rwth-aachen.de }
\end{center}}\bigskip

\begin{longtable}{lp{11cm}}

2010-01 $^\ast$ &Fachgruppe Informatik:      Jahresbericht 2010\\
2010-02 & Daniel Neider, Christof L\"{o}ding:         Learning Visibly One-Counter Automata in Polynomial Time\\
2010-03 & Holger Krahn:         MontiCore: Agile Entwicklung von dom\"{a}nenspezifischen Sprachen im Software-Engineering\\
2010-04 & Ren\'{e} W\"{o}rzberger:         Management dynamischer Gesch\"{a}ftsprozesse auf Basis statischer Prozessmanagementsysteme\\
2010-05 & Daniel Retkowitz:         Softwareunterst\"{u}tzung f\"{u}r adaptive eHome-Systeme\\
2010-06 & Taolue Chen, Tingting Han, Joost-Pieter Katoen, Alexandru Mereacre:         Computing maximum reachability probabilities in Markovian timed automata\\
2010-07 & George B. Mertzios:         A New Intersection Model for Multitolerance Graphs, Hierarchy, and Efficient Algorithms\\
2010-08 & Carsten Otto, Marc Brockschmidt, Christian von Essen, J\"{u}rgen Giesl:         Automated Termination Analysis of Java Bytecode by Term Rewriting\\
2010-09 & George B. Mertzios, Shmuel Zaks:         The Structure of the Intersection of Tolerance and Cocomparability Graphs\\
2010-10 & Peter Schneider-Kamp, J\"{u}rgen Giesl, Thomas Str\"{o}der, Alexander Serebrenik, Ren\'{e} Thiemann:         Automated Termination Analysis for Logic Programs with Cut\\
2010-11 & Martin Zimmermann:         Parametric LTL Games\\
2010-12 & Thomas Str\"{o}der, Peter Schneider-Kamp, J\"{u}rgen Giesl:         Dependency Triples for Improving Termination Analysis of Logic Programs with Cut\\
2010-13 & Ashraf Armoush:         Design Patterns for Safety-Critical Embedded Systems\\
2010-14 & Michael Codish, Carsten Fuhs, J\"{u}rgen Giesl, Peter Schneider-Kamp:         Lazy Abstraction for Size-Change Termination\\
2010-15 & Marc Brockschmidt, Carsten Otto, Christian von Essen, J\"{u}rgen Giesl:         Termination Graphs for Java Bytecode\\
2010-16 & Christian Berger:         Automating Acceptance Tests for Sensor- and Actuator-based Systems on the Example of Autonomous Vehicles\\
2010-17 & Hans Gr\"{o}nniger:         Systemmodell-basierte Definition objektbasierter Modellierungssprachen mit semantischen Variationspunkten\\
2010-18 & Ibrahim Arma\c{c}:         Personalisierte eHomes: Mobilit\"{a}t, Privatsph\"{a}re und Sicherheit\\
2010-19 & Felix Reidl:         Experimental Evaluation of an Independent Set Algorithm\\
2010-20 & Wladimir Fridman, Christof L\"{o}ding, Martin Zimmermann:         Degrees of Lookahead in Context-free Infinite Games\\
2011-01 $^\ast$ &Fachgruppe Informatik:      Jahresbericht 2011\\
2011-02 & Marc Brockschmidt, Carsten Otto, J\"{u}rgen Giesl:         Modular Termination Proofs of Recursive Java Bytecode Programs by Term Rewriting\\
2011-03 & Lars Noschinski, Fabian Emmes, J\"{u}rgen Giesl:         A Dependency Pair Framework for Innermost Complexity Analysis of Term Rewrite Systems\\
2011-04 & Christina Jansen, Jonathan Heinen, Joost-Pieter Katoen, Thomas Noll:         A Local Greibach Normal Form for Hyperedge Replacement Grammars\\
2011-06 & Johannes Lotz, Klaus Leppkes, and Uwe Naumann:         dco/c++ - Derivative Code by Overloading in C++\\
2011-07 & Shahar Maoz, Jan Oliver Ringert, Bernhard Rumpe:         An Operational Semantics for Activity Diagrams using SMV\\
2011-08 & Thomas Str\"{o}der, Fabian Emmes, Peter Schneider-Kamp, J\"{u}rgen Giesl, Carsten Fuhs:         A Linear Operational Semantics for Termination and Complexity Analysis of ISO Prolog\\
2011-09 & Markus Beckers, Johannes Lotz, Viktor Mosenkis, Uwe Naumann (Editors):         Fifth SIAM Workshop on Combinatorial Scientific Computing\\
2011-10 & Markus Beckers, Viktor Mosenkis, Michael Maier, Uwe Naumann:         Adjoint Subgradient Calculation for McCormick Relaxations\\
2011-11 & Nils Jansen, Erika Ábrah\'{a}m, Jens Katelaan, Ralf Wimmer, Joost-Pieter Katoen, Bernd Becker:         Hierarchical Counterexamples for Discrete-Time Markov Chains\\
2011-12 & Ingo Felscher, Wolfgang Thomas:         On Compositional Failure Detection in Structured Transition Systems\\
2011-13 & Michael F\"{o}rster, Uwe Naumann, Jean Utke:         Toward Adjoint OpenMP\\
2011-14 & Daniel Neider, Roman Rabinovich, Martin Zimmermann:         Solving Muller Games via Safety Games\\
2011-16 & Niloofar Safiran, Uwe Naumann:         Toward Adjoint OpenFOAM\\
2011-17 & Carsten Fuhs:         SAT Encodings: From Constraint-Based Termination Analysis to Circuit Synthesis
\\
2011-18 & Kamal Barakat:         Introducing Timers to pi-Calculus\\
2011-19 & Marc Brockschmidt, Thomas Str\"{o}der, Carsten Otto, J\"{u}rgen Giesl:         Automated Detection of Non-Termination and NullPointerExceptions for Java Bytecode\\
2011-24 & Callum Corbett, Uwe Naumann, Alexander Mitsos:         Demonstration of a Branch-and-Bound Algorithm for Global Optimization using McCormick Relaxations\\
2011-25 & Callum Corbett, Michael Maier, Markus Beckers, Uwe Naumann, Amin Ghobeity, Alexander Mitsos:         Compiler-Generated Subgradient Code for McCormick Relaxations\\
2011-26 & Hongfei Fu:         The Complexity of Deciding a Behavioural Pseudometric on Probabilistic Automata\\
2012-01 & Fachgruppe Informatik:         Annual Report 2012\\
2012-02 & Thomas Heer:         Controlling Development Processes\\
2012-03 & Arne Haber, Jan Oliver Ringert, Bernhard Rumpe:         MontiArc - Architectural Modeling of Interactive Distributed and Cyber-Physical Systems\\
2012-04 & Marcus Gelderie:         Strategy Machines and their Complexity\\
2012-05 & Thomas Str\"{o}der, Fabian Emmes, J\"{u}rgen Giesl, Peter Schneider-Kamp, and Carsten Fuhs:         Automated Complexity Analysis for Prolog by Term Rewriting\\
2012-06 & Marc Brockschmidt, Richard Musiol, Carsten Otto, J\"{u}rgen Giesl:         Automated Termination Proofs for Java Programs with Cyclic Data\\
2012-07 & Andr\'{e} Egners, Bj\"{o}rn Marschollek, and Ulrike Meyer:         Hackers in Your Pocket: A Survey of Smartphone Security Across Platforms\\
2012-08 & Hongfei Fu:         Computing Game Metrics on Markov Decision Processes\\
2012-09 & Dennis Guck, Tingting Han, Joost-Pieter Katoen, and Martin R. Neuh\"{a}u\ss{}er:         Quantitative Timed Analysis of Interactive Markov Chains\\
2012-10 & Uwe Naumann and Johannes Lotz:         Algorithmic Differentiation of Numerical Methods: Tangent-Linear and Adjoint Direct Solvers for Systems of Linear Equations\\
2012-12 & J\"{u}rgen Giesl, Thomas Str\"{o}der, Peter Schneider-Kamp, Fabian Emmes, and Carsten Fuhs:         Symbolic Evaluation Graphs and Term Rewriting --- A General Methodology for Analyzing Logic Programs\\
2012-15 & Uwe Naumann, Johannes Lotz, Klaus Leppkes, and Markus Towara:         Algorithmic Differentiation of Numerical Methods: Tangent-Linear and Adjoint Solvers for Systems of Nonlinear Equations\\
2012-16 & Georg Neugebauer and Ulrike Meyer:         SMC-MuSe: A Framework for Secure Multi-Party Computation on MultiSets\\
2012-17 & Viet Yen Nguyen:         Trustworthy Spacecraft Design Using Formal Methods\\
2013-01 $^\ast$ &Fachgruppe Informatik:      Annual Report 2013\\
2013-02 & Michael Reke:         Modellbasierte Entwicklung automobiler Steuerungssysteme in Klein- und mittelst\"{a}ndischen Unternehmen\\
2013-03 & Markus Towara and Uwe Naumann:         A Discrete Adjoint Model for OpenFOAM\\
2013-04 & Max Sagebaum, Nicolas R. Gauger, Uwe Naumann, Johannes Lotz, and Klaus Leppkes:         Algorithmic Differentiation of a Complex C++ Code with Underlying Libraries\\
2013-05 & Andreas Rausch and Marc Sihling:         Software \& Systems Engineering Essentials 2013\\
2013-06 & Marc Brockschmidt, Byron Cook, and Carsten Fuhs:         Better termination proving through cooperation\\
2013-07 & Andr\'{e} Stollenwerk:         Ein modellbasiertes Sicherheitskonzept f\"{u}r die extrakorporale Lungenunterst\"{u}tzung\\
2013-08 & Sebastian Junges, Ulrich Loup, Florian Corzilius and Erika Ábrah\'{a}m:         On Gr\"{o}bner Bases in the Context of Satisfiability-Modulo-Theories Solving over the Real Numbers\\
2013-10 & Joost-Pieter Katoen, Thomas Noll, Thomas Santen, Dirk Seifert, and Hao Wu:         Performance Analysis of Computing Servers using Stochastic Petri Nets and Markov Automata\\
2013-12 & Marc Brockschmidt, Fabian Emmes, Stephan Falke, Carsten Fuhs, and J\"{u}rgen Giesl:         Alternating Runtime and Size Complexity Analysis of Integer Programs\\
2013-13 & Michael Eggert, Roger H\"{a}u\ss{}ling, Martin Henze, Lars Hermerschmidt, Ren\'{e} Hummen, Daniel Kerpen, Antonio Navarro P\'{e}rez, Bernhard Rumpe, Dirk Thi\ss{}en, and Klaus Wehrle:         SensorCloud: Towards the Interdisciplinary Development of a Trustworthy Platform for Globally Interconnected Sensors and Actuators\\

\end{longtable}
\bigskip

\noindent
{\small $^\ast$ These reports are only available as a printed version.\\
  Please contact \email{biblio@informatik.rwth-aachen.de} to obtain
  copies.}

\end{document}